# Topological Insulator Laser Using Valley-Hall Photonic Crystals

*Yongkang Gong[1], Stephan Wong[1], Anthony J. Bennett[2], Diana L. Huffaker[1], and Sang Soon Oh[1],\**

*E-mail: OhS2@cardiff.ac.uk

[1]School of Physics and Astronomy, Cardiff University, Cardiff, CF24 3AA, United Kingdom.

[2]School of Engineering, Cardiff University, Cardiff, CF24 3AA, United Kingdom.

Topological photonics has recently been proved a robust framework for manipulating light. Active topological photonic systems, in particular, enable richer fundamental physics by employing nonlinear light-matter interactions, thereby opening a new landscape for applications such as topological lasing. Here we report an all-dielectric topological insulator laser scheme based on semiconductor cavities formed by topologically distinct Kagome photonic crystals. The proposed planar semiconductor Kagome lattice allows broadband edge states below the light line due to photonic valley hall effect in telecommunication region, which provides a new route to retrieve nontrivial photonic topology and to develop integrated topological systems for robust light generation and transport.

Transplanting the concept of topological phase transitions in fermionic systems to photonics has recently attracted enormous interest[1, 2]. In contrast to conventional photonic defect states that are sensitive to perturbations, edge states from two topologically distinct regions in photonic topological insulators (PTIs) are robust against local perturbations and immune to back scattering, which could lead to intriguing and unexpected photonic devices and functionalities for robust electromagnetic wave transport and processing[3-6]. Implementations of PTIs vary considerably and several schemes have been developed. For example, photonic analogue of a quantum Hall topological insulator was first achieved in the microwave regime using gyromagnetic materials with a strong magnetic field applied to break the time-reversal symmetry[7]. Later, a number of proposals have been put forward to realize photonic topological transport free of external fields by temporal modulation of photonic crystals to mimic time-reversal-symmetry breaking[8-11]. A subwavelength-scale PTI approach via pseudo-time-reversal symmetry in all-dielectric photonic crystals was recently proposed[12] and experimentally verified from visible wavelengths to microwave regime[13-17]. Another elegant subwavelength-scale nontrivial topology strategy has been introduced by exploiting the optical valley Hall effect to break spatial-inversion symmetry to access the opposite Berry curvature near Brillouin zone corners[18-26], which opens avenues to on-chip photonic devices for robust topologically protected light manipulation.

In addition to above passive PTIs researches, a considerable effort has been made towards the study of non-Hermitian PTIs by engaging edge states with optical nonlinearity to enable topological lasing. In contrast to the conventional laser technologies, which are generally sensitive to cavity deformations caused by fabrication imperfections and fluctuations, topological insulator laser cavities are potentially immune to certain cavity defects with higher lasing efficiency attributed to the unique characteristics of nontrivial edge states. Topological insulator laser was first experimentally reported in magneto-optical photonic crystals pumped by a static magnetic field to break time-reversal symmetry[27]. Although this approach allows nonreciprocal lasing from topological cavities of arbitrary geometries, it produces narrow topological bandgap due to weak magneto-optic effect in optical regime. Non-magnetic topologically protected edge-mode lasing was later proposed and implemented by nontrivial semiconductor ring-resonator arrays, and high efficiency single mode lasing that is robust to cavity defects/disorders was reported[28, 29]. The one-dimensional Su-Schrieffer-Heeger (SSH) hamiltonian model is another popular approach to generate edge states and various types of SSH lasing devices have been recently demonstrated based on micro-ring resonators[30, 31], electrically injected Fabry-Perot chain[32], semiconductor pillar arrays[33], and photonic crystal nanocavities[34, 35]. Very recently, a topological single mode laser based on bulk topological effect was reported[36].

In this paper, we propose a new kind of all-dielectric photonic topological laser based on Kagome valley-hall photonic-crystals (KVPs) that consist of hexagonal lattice with primitive cells containing three nanoholes in compound semiconductor membrane. Our analytical and numerical investigations demonstrate that the proposed KVPs concept can lift degeneracy at K point with geometrical perturbation and open broad photonic bandgaps. Valley-dependent edge states and topologically robust transport with subwavelength scale confinement are observed at the interface between two perturbed KVPs with different valley Chern numbers. We further construct topological laser cavities with triangular geometry and explore pumping and lasing dynamics of the topological cavities by means of a four-level two-electron model. Our demonstration of lasing from the proposed KVPs scheme could provide opportunities for engineering topological edge states for various passive and active photonic integrated devices and systems.

## Results
**Topological properties of the KVPs.** The proposed all-dielectric topological strategy is based on hole array KVPs on high refractive index InGaAsP membrane, which has hexagonal lattice with primitive cell composed of three nanoholes with identical diameters $D$, as schematically depicted in Fig. 1a. Perturbation to retrieve nontrivial KVPs can be introduced by putting the three nanoholes closer (negative perturbation) or further away (positive perturbation) from each other. When the hole-to-center spacing $d$ equals to $d_0 = \frac{a}{2\times\sqrt{3}}$, the KVPs are typical Kagome photonic crystal with $C_6$ symmetry, featuring a Dirac cone at K and K′ points in the momentum space as indicated by the TE-like band structure (Fig. 1b). When we introduce positive perturbation by varying $d=d_0$ to $d=1.1d_0$ or negative perturbation by varying $d=d_0$ to $d=0.9d_0$, we break the inversion symmetry and reduce the lattice symmetry to $C_3$ symmetry. As a result, the degeneracy at the K (K′) point is lifted and a band gap opens. Although the positively and negatively perturbed KVPs show the same band structure (Figure 1b) due to the geometrical equivalence, they produce 180 degree-rotated Berry curvature profile because the K and K′ points are swapped under the rotation as illustrated in Fig. 1c and 1d, respectively, where non-zero Berry curvature with opposite signs appears at the K and K′ points. We observe that the negatively perturbed KVPs have positive (negative) Berry curvature at K (K′), while the positively perturbed KVPs have negative (positive) Berry curvature at K (K′). As a result, the two types of KVPs have valley Chern numbers with different signs, thereby edge states are guaranteed at the geometric boundary between them. The band structure and Berry curvature results obtained from our tight binding approximation (Fig. S2 and S3 in the Supplementary Information) agree well with the those in Figs. 1b-1d.

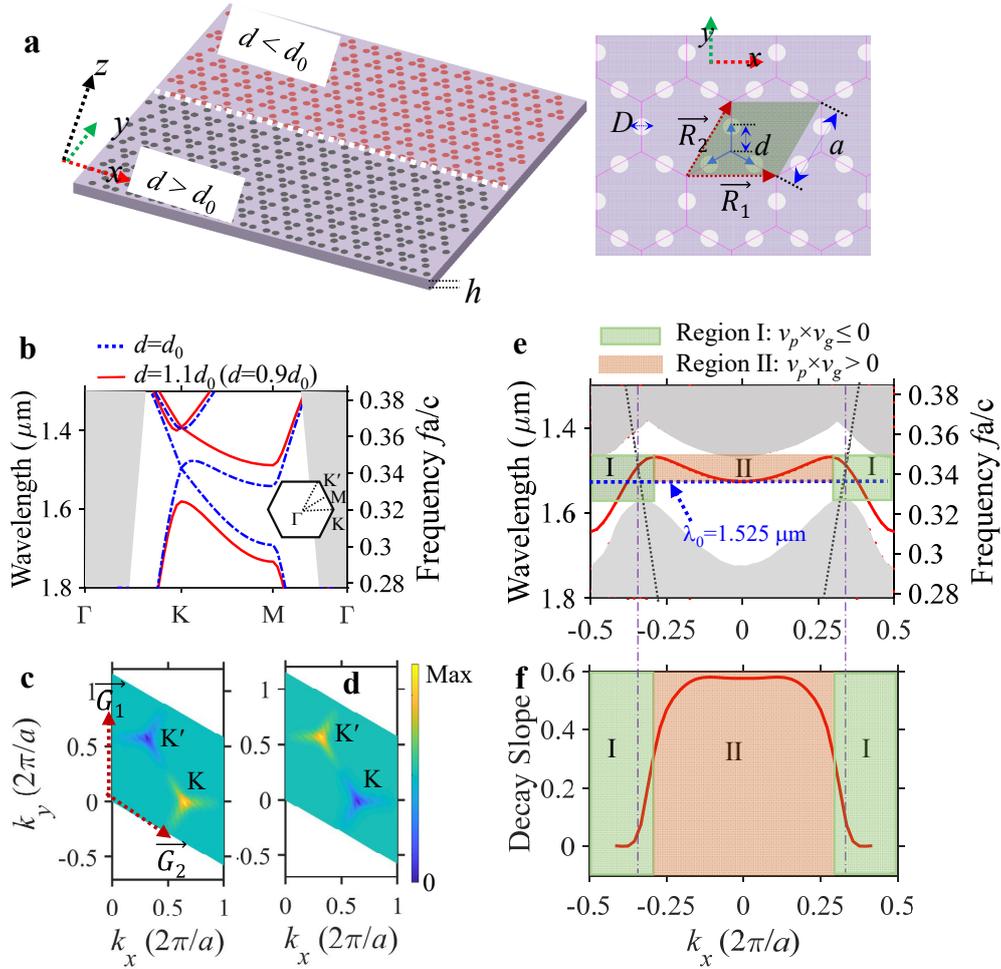

**Fig. 1 Band structure and topology analysis of the proposed all-dielectric KVPs on InGaAsP membrane. a** Schematic diagram of the perturbed KVPs (left) and top view of the KVPs (right) indicating primitive cell that contains three holes with unit vectors of $\vec{R_1}$ and $\vec{R_2}$ and lattice constant length of *a*. The positive and negative perturbation occurs at spacing $d>d_0$ and $d<d_0$ ($d_0 = \frac{a}{2\times\sqrt{3}}$), respectively. **b** Band structure of the unperturbed (blue line) and perturbed (red line) KVPs. The grey region represents light cone. **c, d** Berry curvature of the first band of the negatively and positively perturbed KVPs in the reciprocal primitive unit cell, respectively. $\vec{G_1}$ and $\vec{G_2}$ are the reciprocal lattice vectors. **e** Projected band structure of the topological waveguide in **a**. **f** Decay slope of the edge states of the topological waveguide. In **e** and **f**, the black dashed line represents light lines, the purple perpendicular dashed lines indicate the wavelengths that the light lines and the edge dispersion curve cross, and the blue parallel dotted line shows edge mode wavelength $\lambda_0=1.525$ μm at $k_x=0$. The geometrical parameters in above designs are slab height $h=170$ nm, lattice length $a=500$ nm, and hole diameter $D=150$ nm. The unperturbed KVPs has $d=d_0=144$ nm, while the positively and negatively perturbed KVPs have $d=1.1d_0$ and $d=0.9d_0$, respectively. The refractive index of InGaAsP material is 3.3[37].

**Topological light routing.** Thanks to the bulk-edge correspondence, we can achieve topological edge states at the boundary between the KVCs with positive and negative perturbation as shown in Fig. 1a. Figure 1e reveals that edge states with wavelength range of 1468 nm-1578 nm are generated within photonic bandgap. Moreover, edge states at wavelengths of 1500 nm-1578 nm are below light line, which guarantees broadband topologically protected light propagation with out-of-plane confinement. It is worth noting that the edge states can be categorized into two regions: region I ($v_p \times v_g \leq 0$) and region II ($v_p \times v_g > 0$), where $v_p$ and $v_g$ are phase velocity and group velocity, respectively. We observe from Fig. 1e that when edge mode wavelength is longer than $\lambda_0$, only single unidirectional edge mode within Region I can be excited at each wavelength in the light propagation direction of +$x$ or -$x$. When the edge mode wavelength is shorter than (or equal to) $\lambda_0$, however, dual unidirectional edge modes (one in Region I and the another in Region II) can be excited at each wavelength in the light propagation direction of +$x$ or -$x$. Figure 1f demonstrates that the edge modes

(especially those below light line) in Region I have smaller decay slope and thus experience lower propagation loss than Region II.

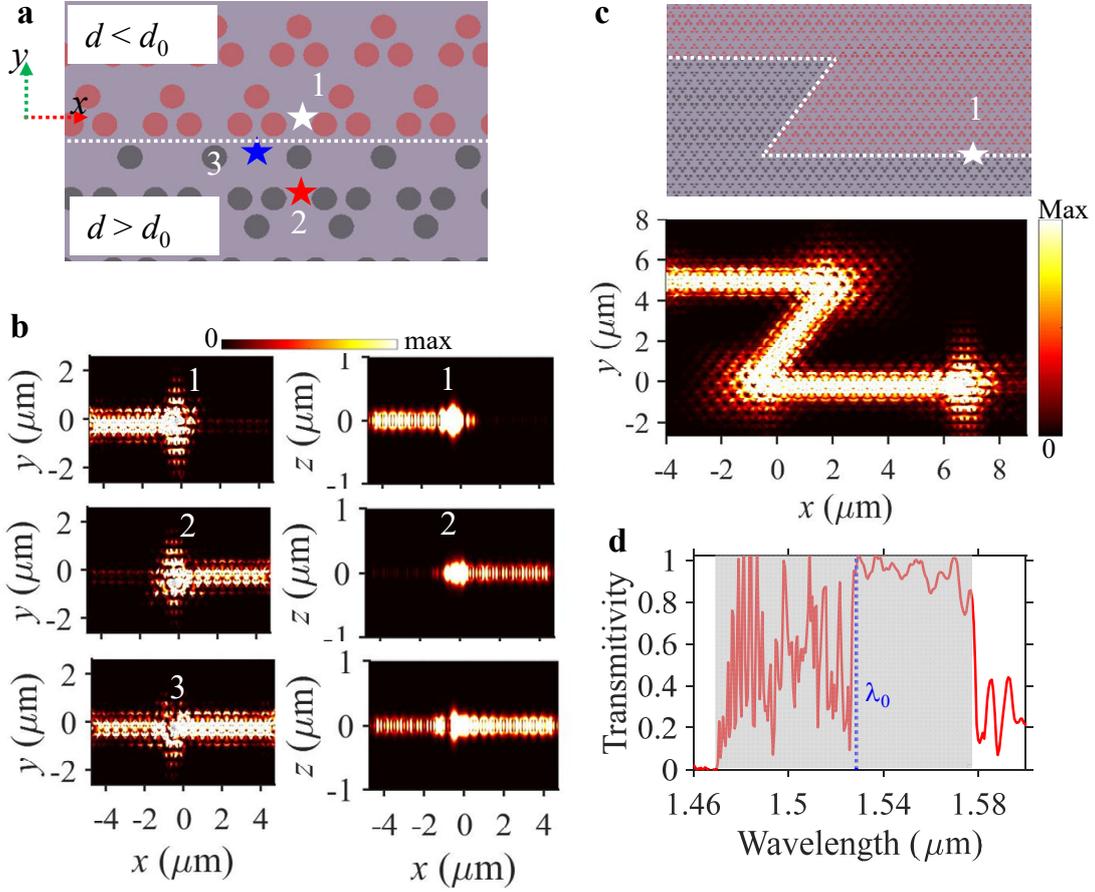

**Fig. 2 Light propagation and field distribution characteristics of the edge states**. **a** Left-handed circular polarized dipoles (marked as star symbols) placed at different positions near the interface between the negatively and the positively perturbed KVPs to excite edge modes and **b** the corresponding field distribution $|E|^2$ of the edge mode at wavelength of 1550 nm in *xy* and *xz* planes. **c** Schematic of a Z-shaped topological waveguide (upper figure) and the corresponding field distribution $|E|^2$ at wavelength of 1550 nm (bottom figure). **d** Transmission spectra of the Z-shaped waveguide obtained by normalizing the Z-shaped waveguide to straight waveguide. The shaded area indicates the wavelength range of the edge states. The white dashed lines in **c** and **d** indicate the boundary between the negatively and the positively perturbed KVPs. The geometric parameters in above designs are the same to those in Fig. 1.

We investigate unidirectional propagation feature of the edge states in KVP waveguide as schematically shown in Fig. 2a. Figure 2b shows that the same sources at different structure positions can excite unidirectional edge states propagating in the opposite direction due to chiral light matter interaction[38], which is also observed in other reported photonic quantum spin Hall[12, 14, 15] and photonic quantum valley Hall systems[25, 26]. When left-handed circularly polarized source excites photons at locations "1" ("2"), negative (positive) chirality is predominant and leftward (rightward) unidirectional light propagation occurs. At location "3", the local electric fields of the edge mode are elliptically polarized and hence the light propagates in both left and right direction[39, 40]. In any case, the light is well confined at the boundary between the two perturbed KVPs in both *xy* and *xz* planes due to the presence of topological edge states below light line. We also evaluate field distribution and spectral properties of topological waveguides with 60-degree bends. Figure 2c and 2d show that the topologically protected light propagates smoothly around sharp bends with high transmission. Spectral oscillation occurs at wavelengths shorter than $\lambda_0$=1.525 μm. It is because edge modes in both Region I and Region II are excited in this case (as predicted in Fig. 1e), and the intensity of the excited modes in the two regions varies significantly with wavelength under the same dipole source excitation.

**Q factor and resonant modes of the KVP cavities.** Based on the developed KVP scheme, we design a topological triangle cavity that consists of the negatively perturbed KVPs inside of the cavity and the positively perturbed KVPs outside of the cavity, as depicted in Fig. 3a. If the cavity is arranged in an opposite manner, i.e., the cavity inside is the positively perturbed KVPs while the cavity outside is the negatively perturbed KVPs, the majority of the edge states lies above the light line (see Fig. S6 in Supplementary Information). The proposed cavity supports two types of resonant modes: ring-resonator modes and FP-resonator modes. For example, when cavity edge length *L* is 14.5 μm, nine ring-resonator modes with frequencies of $f_i$ (*i*=1, 2, …9) and seven FP (Fabry–Pérot) -resonator modes with frequencies of $f_m'$ (*m*=1, 2, …7) appear, as shown in Fig. 3b. The ring-resonator modes arise from excitation of low propagation loss edge modes within Region I (Figs. 1e and 1f), which is evidenced by the $E$-field distributions showing that light propagates smoothly along the whole cavity (Figs. 3d and 3e). In contrast to the ring-resonator modes, the FP-resonator modes experience high reflection from the cavity corners (Figs. 3f and 3g). The FP-resonator modes originate from excitation of lossy edge modes in Region II (Figs. 1e and 1d), which explains why the wavelengths of FP-resonator modes are shorter than $\lambda_o$ as indicated in Fig. 3b.

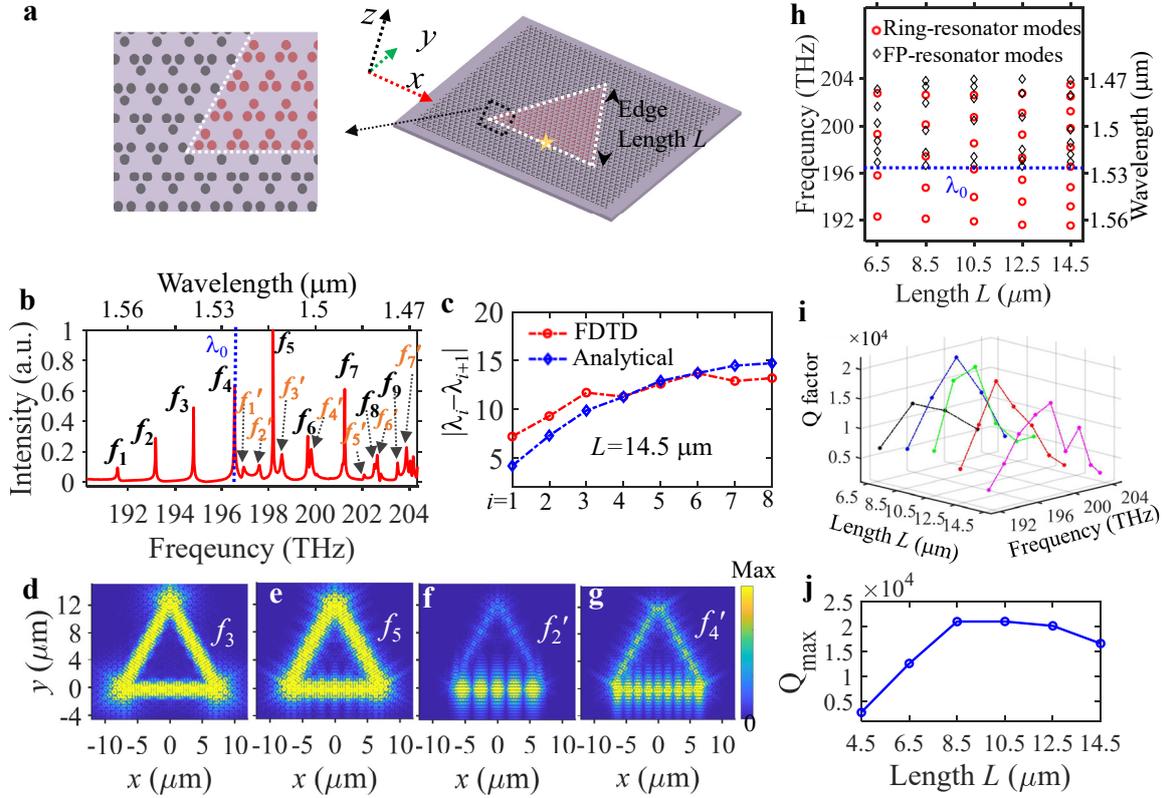

**Fig. 3 Optical properties of the proposed equilateral triangle KVP cavity**. **a** Schematic diagram of the cavity (right) and the zoomed in cavity corner area (left). Red and black areas represent the negatively and positively perturbed KVPs, respectively, and the white dashed line indicate the boundary between them. The star symbol indicates the location where dipole sources are placed to excite cavity modes. **b** Optical spectra of the KVP cavity under cavity edge length of $L$=14.5 μm. **c** Numerically and analytically derived FSR of the ring-resonator modes. **d, e** Field distribution $|E|$ of the topological ring-resonator modes at frequencies of $f_3$=194.8 THz ($\lambda_3$=1539 nm) and $f_5$=198.2 THz ($\lambda_5$=1513 nm) at $Z$=0 μm plane, respectively. **f, g** Field distribution $|E|$ of the FP-resonator modes at frequencies of $f_2'$=197.6 THz and $f_4'$=199.8 THz, respectively. **h** Dependence of the frequencies of the resonant modes on cavity length. **i** Dependence of the ring-resonator modes' Q factor on cavity length. **j** The maximum Q factor ($Q_{max}$) versus cavity length. The geometrical parameters in the simulations are the same to those in Fig. 2.

To further analyze the cavity modes, we perform free-spectral-range (FSR) calculations of the ring-resonator modes based on $\Delta\lambda_{FSR}=\lambda^2/(n_g L_{cavity})$, where $n_g$ is group index and $L_{cavity}$ is the effective length of light propagating in the cavity. Numerical FSR results were obtained from FDTD results using $\Delta\lambda_{FSR}=|\lambda_i-\lambda_{i+1}|$ ($i$=1, 2, …8), where $\lambda_i$ is the wavelength of $i$-th ring resonator modes. Figure 3c illustrates that the value of the FSR of the ring-resonator modes is not constant due to waveguide dispersion (i.e., $n_g$ varies with wavelength). There is deviation between the analytically and the numerically obtained FSR, which arises from the fact that we approximate $L_{cavity}$=3$L$ in the numerical analysis. In fact, $L_{cavity}$ should change with wavelength, since different ring-resonator modes have slightly different light field profile around cavity corners and thus yield different $L_{cavity}$. Furthermore, we study the dependence of the resonant modes' frequencies on the cavity length (Fig. 3h). The frequency spacing of the ring-resonator modes decrease with $L$. We note that the FP-resonator modes always appear at wavelengths shorter than $\lambda_0$=1.525 μm, which prove that the physical origin of FP-resonator modes is excitation of the edge modes in Region II (Fig. 1e and 1d). Figure 3i shows that for the same cavity length, the ring-resonator modes at the frequencies in the sides of the edge states range tend to have smaller Q factors than those for the frequencies in the middle of the edge states range, which is because they are close to bulk modes and have larger lateral losses. We observe that the maximum Q factor reaches 2.1×10$^4$ at wavelength of 1518.7 nm when the cavity edge length $L$=8.5 μm. The proposed cavity has higher Q factor than that of the recently reported topological ring resonator cavity based on photonic quantum spin hall effect[17]. Figure 3j demonstrates that the maximum Q factor depends on cavity length and decreases rapidly with the cavity length when $L<8.5$ μm, which is due to light in the cavity edges starting to couple from each other when the cavity edges become closer. We also note that the maximum Q factor drops when cavity length increases ($L>8.5$ μm) due to larger cavity loss for longer cavity, but it drops slowly because of the propagation loss of the modes below the light line (Figs. 1e and 1f), which differs fundamentally to the spin Hall-type photonic crystal cavities that have larger vertical loss and lower Q factor due to edge states above the light line[17, 41].

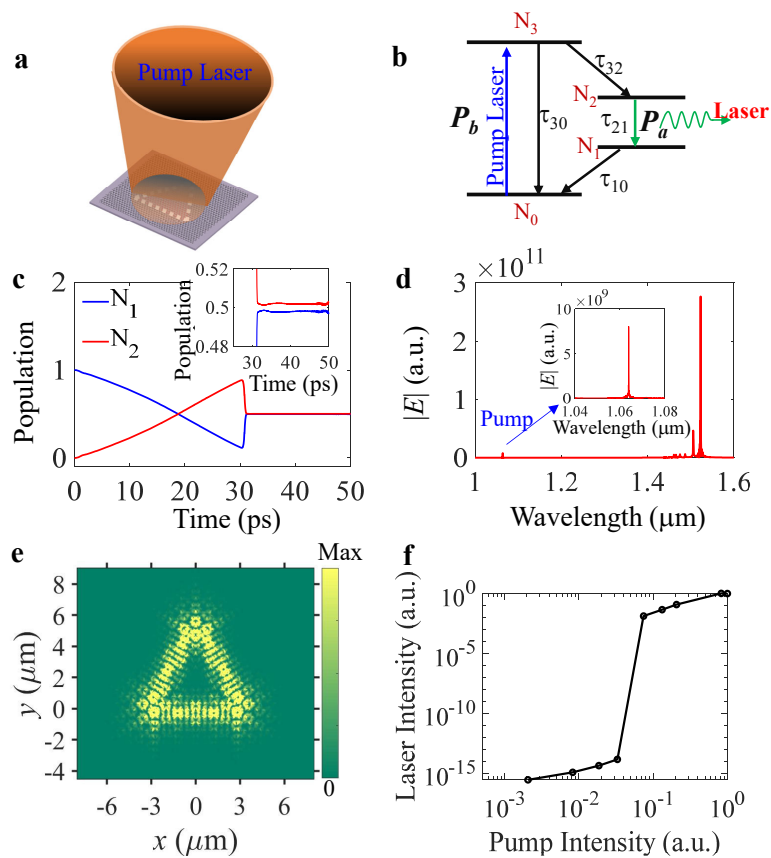

**Fig. 4**. **Lasing characteristics of the proposed triangle cavity KVP lasers**. **a** Schematic of the cavity optically pumped by another laser from the top of it. **b** Schematic diagram of the four-level two-electron model for the KVP lasers. **c** Evolution of electron population density probability with time at level $N_1$ and $N_2$. The inset shows population inversion when the lasing is stabilized. **d** Spectrum of the KVP laser. The inset is the spectrum of the pump laser at wavelength of 1064 nm. **e** Lasing field distribution $|E|$ of the cavity at wavelength of ~1523 nm at $Z=0$ μm plane. **f** Normalized output laser intensity versus normalized pump laser intensity. The geometrical parameters in the simulations are the same to those in Fig. 3.

**Topological laser cavities**. We further study lasing performance of the topological cavity by introducing gain to the InGaAsP material and optically pumping the KVP cavity, as illustrated in Fig. 4a. We utilize a four-level two-electron model to investigate lasing dynamics of the cavity, as described in Fig. 4b. The electron populations of the levels rely on pumping intensity and spontaneous emission decay. We demonstrate that the population inversion of Level 2 relative to Level 1 can be achieved at pump laser intensity of $|E|=3\times10^5$ V/m (Figure 4c). Consequently, lasing at edge state wavelength of ~1523 nm is achieved, as shown from the laser spectrum and the field distribution in Figs. 4d and 4e. Figure 4f displays that the laser intensity experiences a dramatically jump when the power of the pump laser increases to a certain value and then becomes saturated when the pump laser further increases– a typical laser threshold behavior with a clear transition from spontaneous emission to stimulated emission.

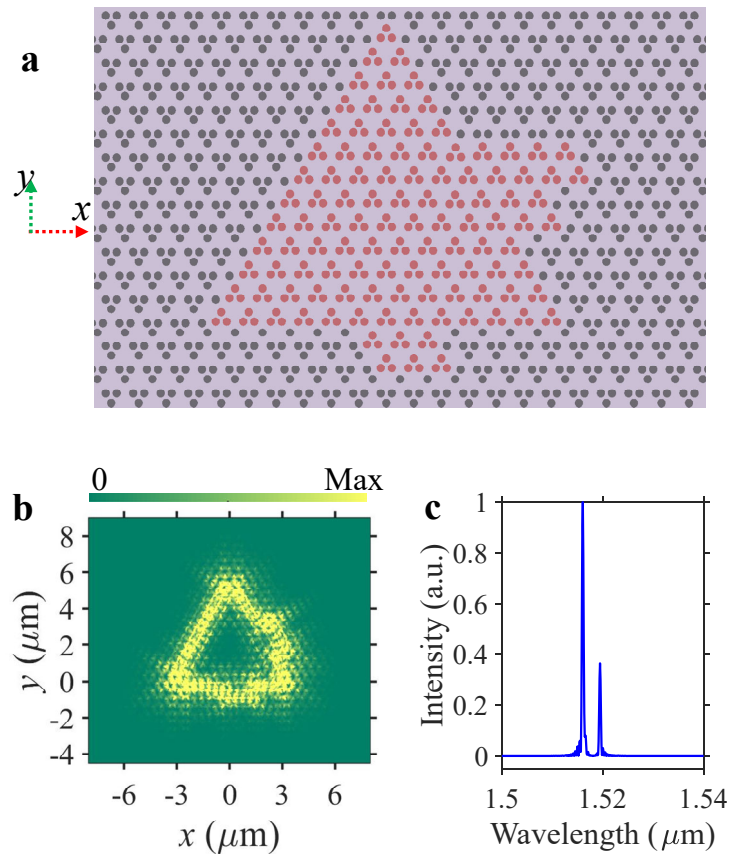

**Fig. 5 Topological insulator laser cavity with complex geometry**. **a** Schematic top-view of the cavity. **b** Field distribution |E| of the laser. **c** Spectrum of the laser cavity. The red and black regions in **a** indicate the negatively and positively perturbed KVPs, respectively.

## Discussion

Discovery of new types of non-trivial photonic topological platforms facilitates the development of novel topological lasers. We have demonstrated a PTI strategy based on all-dielectric hole-type Kagome lattice, which differs fundamentally to the recently reported approaches based on quantum spin Hall effect[27-29, 36]. One advantage of the developed KVPs in the semiconductor membrane platform is that it allows broadband non-trivial generation in telecommunication region below the light line and thus enables robust photonic routing in waveguides even with sharp bends. We explored topologically protected laser cavities based on the KVPs and evaluated the robustness and dynamics of the lasing cavities. Our study helps understanding the interplay between non-Hermiticity and topology, and more importantly provides a new scheme to explore compact all-dielectric topological lasers that can be easily integrated to other passive or active photonic devices by the well-established semiconductor photonic integrated circuit technologies.

## Methods

**Simulation.** The TE-like electromagnetic bandstructure of KVPs based on semiconductor slab were obtained from 3-D MIT Photonic Bands (MPB) software[42]. We retrieve the Berry curvature $F(k)$ from the MPB simulations by scanning wave vector $k$ in the first Brillion zone based on the equation of $F(k) = \nabla_k \times i < u_k | \nabla_k u_k >$, where $u_k$ is the periodic function of the magnetic field $H_z = e^{i\vec{k}\cdot\vec{R}} u_k(\vec{r})$. We characterize the topology of the KVPs by valley Chern number $C_V = C_K - C_{K'}$, where $C_K$ ($C_{K'}$) are Chern number from integration over half of the first Brillouin zone around valley K (K′). We also developed tight binding model and derived the effective Hamiltonian to investigate the topological characteristics of KVP systems (for more details, sections A and B in the Supplementary Information). We optimized the band structure of the edge states by scanning geometrical diameters to have as broad wavelength range of edge modes as possible in the telecommunication regime (Fig. S4 in Supplementary Information).

The projected band structures (Fig. 1e) of the topological waveguides were obtained from full-wave three-dimensional (3-D) Finite-difference time-domain (FDTD) (Lumerical software) calculations of a supercell that consists of both the negatively and positively perturbed KVPs. Since the time domain signal $E(t) \propto e^{i(\omega+i\gamma)t}$ in FDTD method, the envelope of the decaying signal is determined by $\ln(|E(t)|)$ and thus we can retrieve the decay slope of the edge states (Fig. 1f) using equation of $|(\ln|E(t_2)|-\ln|E(t_1)|)|/|t_2-t_1|$, where $\omega$, $\gamma$ and $t_1$ ($t_2$) are angular frequency, decay constant and time, respectively.

The light propagation and field distribution of the edge states in the KVP waveguides were carried out by exciting edge modes with a circular polarized light source placed in specified locations in FDTD modelling (Fig. 2). Transmission spectra of the bended waveguide were implemented by normalizing the waveguide to the straight waveguide with the same geometrical parameters. To get accurate Q factor for the triangular KVP cavities, we placed several dipole sources randomly near one of the cavity edges to excite all the resonant modes and calculated the slope of the envelope of the decaying light signal in the cavities.

We implemented the dynamics of the electron transition and the population intensity modelling of the topological cavities by four-level two-electron model[43] based on 3-D FDTD method (Lumerical software). The four-level atomic system is considered as two coupled dipole oscillators $P_a$ (formed by Level 1 and Level 2) and $P_b$ (formed by Level 0 and Level 3) with angular frequency of $\omega_a$ and $\omega_b$, and dephasing rate $\gamma_a$ and $\gamma_b$ (Fig. 4b), respectively. $N_i$ is electron population density probability in Level $i$ and $\tau_{ij}$ ($i,j=0,1,2,3$) is the decay time constant between levels $i$ and $j$. The parameters for the gain material InGaAsP in the model (Fig. 4 and 5) are $\omega_a$=1.26×10$^{15}$ Hz and $\omega_b$=1.77×10$^{15}$ Hz, $\gamma_a$=1.68×10$^{14}$ Hz, $\gamma_b$=1×10$^{12}$ Hz, $\tau_{30}$=$\tau_{21}$=3×10$^-$

$^{10}$ s, $\tau_{32}=\tau_{10}=1\times10^{-13}$ s[43,44]. We used laser at wavelength of 1064 nm to pump the KVP cavities to excite topological laser modes in telecommunication region.

# Supplementary Information

**Topological Insulator Laser Using Valley-Hall Photonic Crystals**

*Yongkang Gong, Stephan Wong, Anthony J. Bennett, Diana L. Huffaker and Sang Soon Oh\**

*Email: OhS2@cardiff.ac.uk

## A. Derivation of Effective Hamiltonian

In this section, we develop a tight binding model for the proposed KVPs. We calculate the field distribution $|H_z|$ of the perturbed KVPs using MPB software. Figure S1 shows that electromagnetic mode around a hole couples to the other two holes in the same primitive cell (i.e., intracell coupling) and couples to the holes in the nearest neighboring primitive cells as well (i.e., intercell coupling). For the negatively perturbed KVPs, the intracell coupling is stronger than the intercell coupling (Figure S1a), which is the opposite to the case of the positively perturbed KVPs (Figure S1b). Therefore, we can consider our KVPs as 2-D tight binding system with each cell having three sites and consider both the intracell and the intercell coupling, as schematically shown in Figure S2(a). Under the tight binding approximation, Hamiltonian of the KVPs can be given by the following

$$\widehat{H}|\psi\rangle = E|\psi\rangle, \tag{S1}$$

where $E$ is eigenvalue and $\psi$ is the eigenvector. A state that sits on a site $\sigma$ ( $\sigma = A, B, C$ ) in the $(m, n)$ unit cell of our system is denoted by $|m, n, \sigma\rangle$. Due to translation symmetry of the system, any state $|\psi_k\rangle$ can be written as

$$|\psi_k\rangle = \sum_{m,n} e^{ik \cdot R}(c_A|m, n, A\rangle + c_B|m, n, B\rangle + c_C|m, n, C\rangle) \tag{S2}$$

Here, $R = m\vec{R_1} + n\vec{R_2}$ is the translation lattice vector and $C_\sigma$ is a constant to normalize the state $|\psi_k\rangle$.

With the nearest-neighbour approximation, the Hamiltonian of the whole lattice can be constructed by considering both intracell and intercell hopping as

$$\widehat{H} = \widehat{H}_{intra} + \widehat{H}_{inter}, \tag{S3}$$

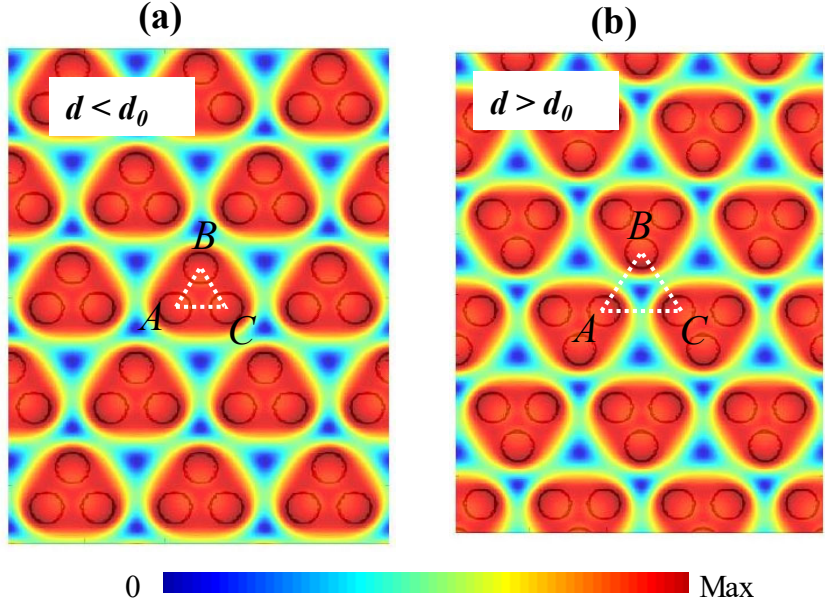

**Figure S1**. Field distribution $|H_z|$ of a) the negatively and b) positively perturbed 3-D KVPs at K point in the first band at $z=0$ μm plane. Labels "*A*", "*B*" and "*C*" indicate the three holes in a primitive cell. The black circles represent the outline of holes. The parameters in the calculation are the same to Figure 1 in the main text.

where the intracell hopping Hamiltonian $\widehat{H}_{intra}$ and intercell hopping Hamiltonian $\widehat{H}_{inter}$ equal to

$$\widehat{H}_{intra} = -v \sum_{m,n} e^{ik \cdot R} (|m,n,B\rangle\langle m,n,A| + |m,n,C\rangle\langle m,n,B| + |m,n,A\rangle\langle m,n,C| + h.c.), \tag{S4}$$

$$\widehat{H}_{inter} = -w \sum_{m,n} e^{ik \cdot R} (|m,n+1,B\rangle\langle m,n,A| + |m+1,n,C\rangle\langle m,n,B| + |m+1,n-1,A\rangle\langle m,n,C| + h.c.). \tag{S5}$$

Here, $k$ is the wave vector, $v$ ($w$) are the intracell (intercell) hopping strength, and *h.c.* denotes the Hermitian conjugate of the previous three states in equations S4 and S5. The eigenvalue equation can be reduced to the effective eigenvalue problem:

$$\widehat{H}_{eff}(k)|\phi\rangle = \varepsilon(k)|\phi\rangle, \tag{S6}$$

where $\varepsilon$ is energy eigenvalue, $|\phi\rangle = (\phi_A, \phi_B, \phi_C)$ is the corresponding three-component

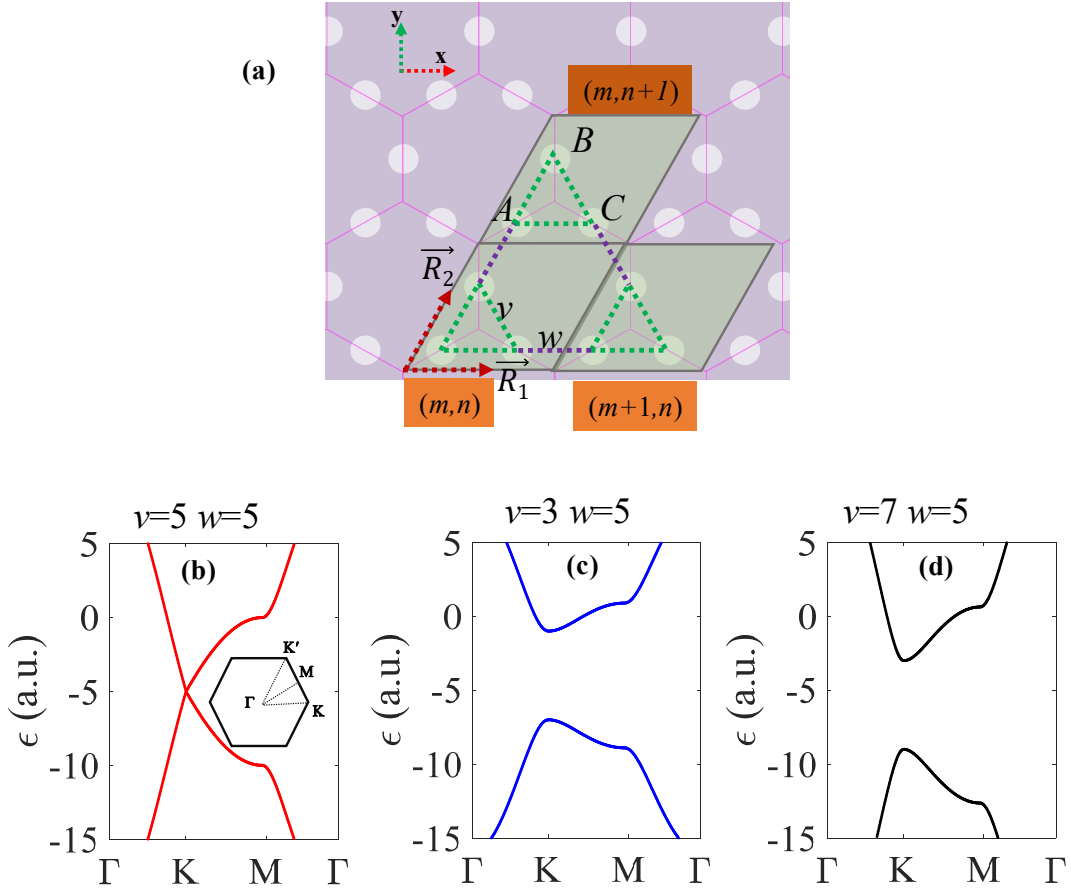

**Figure S2**. Tight binding approximation for effective Hamiltonian derivation. a) Schematic illustration of the tight binding model for the system under study. The structure has triangular lattice with unit cell composed of three sites $A$, $B$ and $C$. Lattice unit vectors are $\vec{R_1}$ and $\vec{R_2}$. $v$ represents the intracell hopping strength of the three sites and $w$ labels intercell hopping strength between two neighboring unit cells. $(m,n)$ represents the location of unit cells. b)-d) Band structure of the unperturbed, negatively perturbed and positively perturbed KVPs, respectively. Bandgap opening occurs in the vicinity of K point when KVPs are perturbed, which agrees well with the results in Figure 1b in the main text.

eigenvector, and $\widehat{H}_{eff}$ is the effective Hamiltonian that reads

$$\hat{H}_{eff}(\vec{k}) = \begin{pmatrix} 0 & -\left(v + we^{i\vec{k}\cdot\vec{R_2}}\right) & -\left(v + we^{-i\vec{k}\cdot(\vec{R_1}-\vec{R_2})}\right) \\ -\left(v + we^{-i\vec{k}\cdot\vec{R_2}}\right) & 0 & -\left(v + we^{-i\vec{k}\cdot\vec{R_1}}\right) \\ -\left(v + we^{i\vec{k}\cdot(\vec{R_1}-\vec{R_2})}\right) & -\left(v + we^{i\vec{k}\cdot\vec{R_1}}\right) & 0 \end{pmatrix}. \quad (S7)$$

By solving the effective Hamiltonian system based on Eqs. S6 and S7, we can obtain the band structure and the Berry curvature of the proposed KVPs.

We calculate the eigenvalue of the developed tight binding system by solving Equation S6. We note from Figure S2b that there is a linear degeneracy at the K point when the Kagome structure is unperturbed (i.e., $v=w$). The degeneracy can be lifted in both the negatively perturbed (i.e., $v>w$) and positively perturbed (i.e., $v<w$) KVPs and bandgap appears at the vicinity of K point, as shown in Figures S2(a) and S2(b). The results agree with the band structure analysis in Figure 2b in the main text, which validates the effective Hamiltonian we derived.

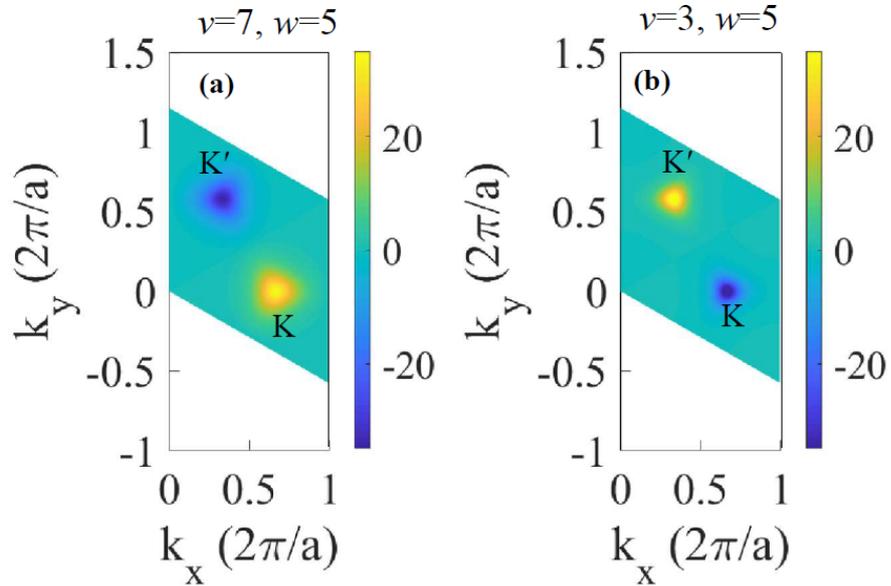

**Figure S3**. Berry curvature of a) the negatively and b) positively perturbed KVPs. The Berry curvature is analytically obtained from Equation S8.

**B. Berry Curvature and Valley Chern number**

Based on the effective Hamiltonian derived in the previous section, we can obtain the Berry curvature and the Chern number. One can analytically derive the Berry curvature $F(k)$ for eigenvalue $\varepsilon_m(k)$ corresponding to an eigenvector $|\Phi_m\rangle$ at a certain $k$ point: [1]

$$F(k) = -\text{Im}\left(\sum_{m \neq n} \frac{\langle\phi_{n,k}|(\nabla_k \hat{H}_{eff})|\phi_{m,k}\rangle \times \langle\phi_{m,k}|(\nabla_k \hat{H}_{eff})|\phi_{n,k}\rangle}{(\varepsilon_n(k)-\varepsilon_m(k))^2}\right). \tag{S8}$$

Here, Im means imaginary part of the equation. Based on Equation S8, we calculate the Berry curvature of the first band of the perturbed KVPs over the primitive unit cell (Figure S3). We notice that the Berry curvature has opposite sign at the K and K' points: the negatively perturbed KVPs have negative Berry curvature at K' and positive Berry curvature at K, while the positively perturbed KVPs have positive Berry curvature at K' and negative Berry curvature at K. The results are in consistent with the calculated Berry curvature in Figures 1d and 1e in the main text.

Based on the obtained Berry curvature we can characterize the topology of the KVP systems by valley Chern number $C_V = C_K - C_{K'}$, [2, 3] where $C_K$ ($C_{K'}$) are Chern number from integration of Berry curvature around valley K (K′). Apparently, the negatively perturbed KVPs have negative $C_{K'}$ and positive $C_K$ due to negative (positive) Berry curvature around K (K′), and thereby possess of negative valley Chern number (i.e., $C_V<0$). On the contrary, the positively perturbed KVPs have positive valley Chern number (i.e., $C_V>0$). Therefore, the positively and negatively perturbed KVPs have different topology, and topological edge states are guaranteed at their boundary as shown in Figure 1a in the main text.

## C. Optimization of KVPs

We optimize the edge states of the proposed 3-D KVPs by scanning slab height $h$, lattice length $a$, and hole diameter $D$, so that the central wavelength of edge states $\lambda_0$ lies in telecommunication region and wavelength range of edge states $\Delta\lambda$ is as broad as possible. Figure S4 demonstrates that $\lambda_0$ increase with slab thickness and lattice and decreases with hole diameter.

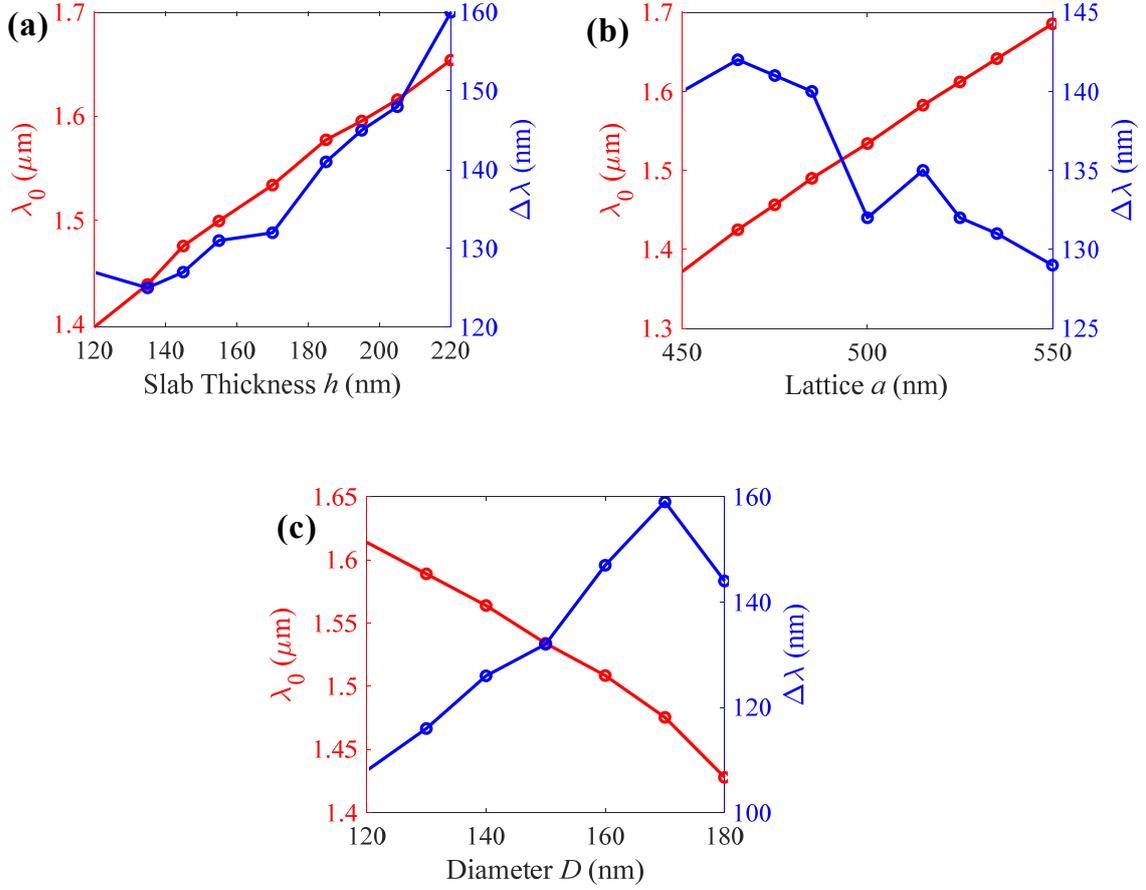

**Figure S4**. Optimization of the proposed KVPs edge states. a) Dependence of $\lambda_0$ and $\Delta\lambda$ on structure slab thickness. The other parameters are $a$=500 nm and $D$=150 nm. b) Dependence of $\lambda_0$ and $\Delta\lambda$ on structure lattice length. The other parameters are $h$=170 nm and $D$=150 nm. c) Dependence of $\lambda_0$ and $\Delta\lambda$ on hole diameter. The other parameters are $h$=170 nm, and $a$=500 nm.

## D. Band structure and Field distribution of KVPs

We plot field distribution $|E|^2$ of light propagation in the Z-bended waveguide (see Figure 2b in the main text) at edge mode wavelength of 1550 nm in both *xy* and *xz* planes. Figure S5 illustrates that light propagates smoothly around the sharp corners and is well confined at the out-of-plane because of the edge modes being below light line.

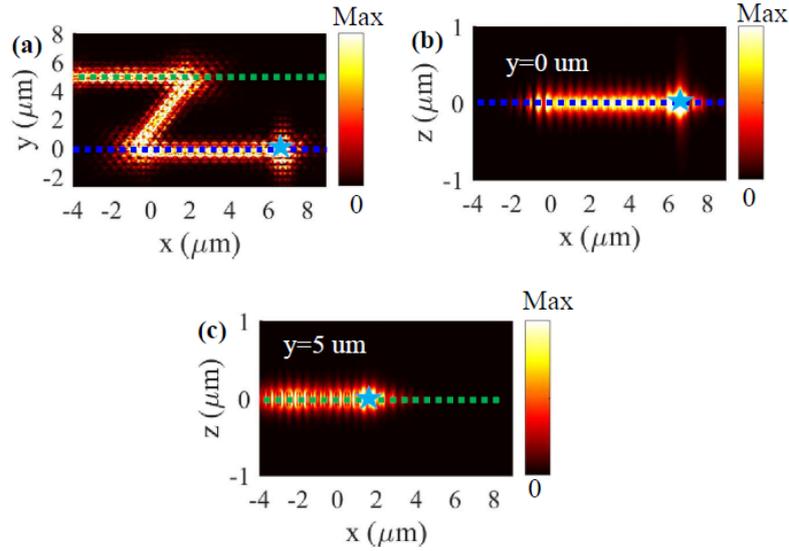

**Figure S5.** a) Field distribution $|E|^2$ of light propagation in Z-bended waveguide in *xy* plane at wavelength of 1550 nm. The star symbol indicates circular polarized source exciting unidirectional light propagation to the left of the waveguide. b)-c) Field distribution $|E|^2$ in the *xz* plane at the locations indicated by the blue and green dashed plane marked in a), respectively.

When we alter the topological edge by arranging the positively and negatively perturbed KVPs in the opposite way to Figure 1a in the main text while maintaining geometrical parameters, as depicted in Figure S6a, although we can still obtain broad edge modes within band gap, the majority of edge modes become lying above the light line (Figure S6(b)).

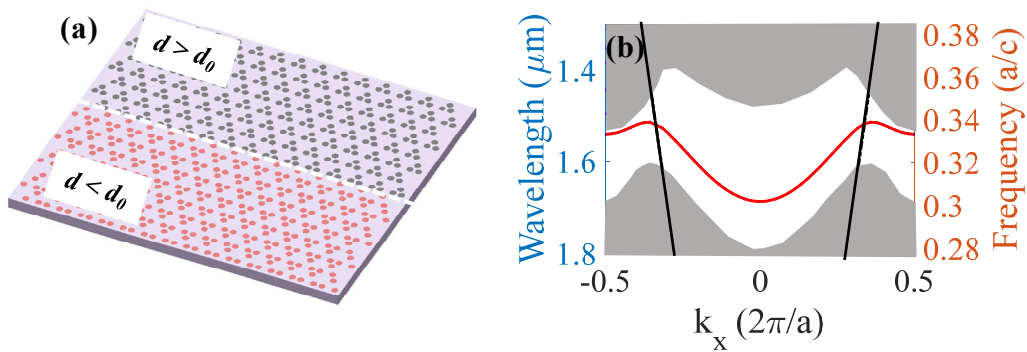

**Figure S6.** a) Schematic of a topological edge composed of positively and negatively perturbed KVPs. b) Band structure of the topological edge. The geometrical parameters used in this calculation are same to that in Figure 1e in the main text.